# Coordinate: Probabilistic Forecasting of Presence and Availability

Eric Horvitz    Paul Koch    Carl M. Kadie    Andy Jacobs

Microsoft Research
One Microsoft Way
Redmond, WA 98052 USA
{horvitz, paulkoch, carlk, andyj}@microsoft.com

**Abstract**

We present methods employed in COORDINATE, a prototype service that supports collaboration and communication by learning predictive models that provide forecasts of users' presence and availability. We describe how data is collected about user activity and proximity from multiple devices, in addition to analysis of the content of users' calendars, the time of day, and day of week. We review applications of presence forecasting embedded in the PRIORITIES application and then present details of the COORDINATE service that was informed by the earlier efforts.

## 1 Introduction

Despite the common use of online calendar systems for storing reminders and creating contracts with others about meeting times and locations, a great deal of collaboration is based on opportunistic communication arranged under uncertainty. Such informal coordination often hinges on peoples' shared intuitions about current and future locations and activities of friends and colleagues. Even with the use of online group calendar systems, people may be challenged with understanding how available others are for various kinds of collaborations [2,10].

We describe methods employed in an automated presence and availability forecasting service named COORDINATE. COORDINATE is targeted at supporting real time, *peri*-real time, and longer-term coordination for messaging and collaboration by providing predictions about the current and future states of users to authorized people and applications. States of interest include a user's current and future presence at one or more locations, availability for interruptions, and other situations including meeting status, receipt of communications, and device access.

Presence and availability forecasting extends existing computational tools that support awareness of people via sharing a user's current state (*e.g.*, online presence). Research on user modeling over the last decade has focused largely on applications that center on reasoning about a user's *current* activities, intentions, and goals [1,4,5,7,8]. The difficulty of determining the goals of users has stimulated interest in representing uncertainty with probabilistic user modeling. Uncertainty plays an even more central role in reasoning about the *future* states of people; even perfect knowledge about a user's current activities and intentions does not typically extinguish uncertainty about the future.

COORDINATE learns forecasting models from observational data that can be enhanced with user input. Although the inferences of COORDINATE may be shared directly with users via direct queries or through applications that support planning and coordination, the motivation and main focus of COORDINATE centers on providing forecasts to automated collaboration, communication, and notification services. Research on presence and availability forecasting evolved within the Attentional User Interface (AUI) effort at Microsoft Research [7], in support of the PRIORITIES, NOTIFICATION PLATFORM, and BESTCOM projects.

We shall first describe an initial implementation of presence forecasting in the PRIORITIES email prioritization and message relay system and review features that relied on the forecasts. We describe the approach we used in PRIORITIES to collect and leverage presence data for generating conditional distributions about a user's future presence. Then, we then move to COORDINATE, a networked Bayesian forecasting service that accesses data from multiple devices and that provides predictions about presence and availability in response to queries. We first describe the system's data collection, model construction, and presence-forecasting capabilities and then dive into details about COORDINATE's interruptability and meeting-analysis subsystems. Finally, we describe the relevance of COORDINATE's reasoning to the NOTIFICATION PLATFORM and BESTCOM projects.

## 2 Forecasting Presence and Availability

Popular online instant messaging (IM) and calendar systems have relied primarily on a user's explicit statements or actions to reveal presence and availability to colleagues. Several systems sense the presence of users and use this information for a variety of services. For example, the PRIORITIES messaging system [7] and its commercialized descendant, Outlook Mobile Manager [12], examine the amount of time a desktop system has been idle. If this period exceeds a prespecified amount of time, the system relays reminders, appointment information, and email messages that have been assigned a sufficient level of urgency to a mobile device.

Later versions of PRIORITIES and related research prototypes [7] consider richer notions of presence beyond desktop activity, including information on users' calendars, computer vision analysis, and ambient acoustical analysis, configured to identify nearby conversations. Fusing multiple channels of evidence allows a computing system to identify a user's presence, as well as to make assessments about a user's attention and intentions, even in the absence of explicit interaction with a computer.

Whether presence is determined via explicit user assertions or by automated processes, the focus to date in real-world systems that attempt to share presence has been on determining and sharing with authorized colleagues a user's *current* status. With presence and availability forecasting, we generalize the focus of attention to include methods and machinery that can predict a user's future situations, given multiple evidential clues.

Beyond basic notions of presence at a specific computer or presence on a computer network, several different classes of events may be of value to people and applications seeking to arrange interactions and communications. For example, we may wish to predict when users will return to their offices if they are currently away. We may wish to reason about notions of *availability*. For example, given that a user is working at a desktop system, or at a meeting somewhere, when might they be free to review a notification, or to receive a communication without being significantly disrupted? We also may wish to know when a user will be available to collaborate for some specified period of time. We may be interested in knowing whether someone will be at a meeting that is listed in their online calendar. Alternatively, we may wish to predict when a user will likely next review their email so as to receive an important message, or next have access to a full desktop computing system for reviewing a document that is being developed collaboratively or to engage in a videoconference. Although queries about a person's future state may vary in sophistication, the methods for making these kinds of predictions are similar. The COORDINATE service was built to handle such queries, and to elucidate core challenges and research issues with forecasting a user's future situations.

### 2.1 Presence Forecasting in Priorities

We shall first describe an early implementation of presence forecasting that was used to endow PRIORITIES with the ability to store and provide probability distributions over the period of a user's absence based on logs of activity on or near a single primary computing device.

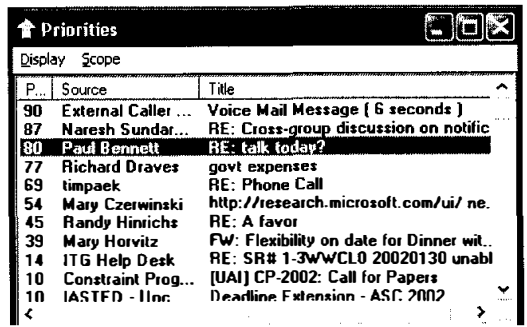

Figure 1. PRIORITIES client displaying unread messages sorted by urgencies and scoped by time. The message received most recently is highlighted by the system.

The PRIORITIES project has centered on the use of classifiers to learn the urgencies of email messages from multiple distinctions gleaned from the header and body of the messages, and the coupling of these inferred urgencies with context-sensitive desktop and mobile alerting policies. The initial PRIORITIES system was distributed for internal use and testing at Microsoft Corporation in 1998 and the prototype system is still in use, even after the system's core functionality was commercialized as the Outlook Mobile Manager.

As background, the current system employs a Support Vector Machine (SVM) classifier with probabilistic output [11] to learn to assign a probability distribution over message urgencies. Training sets are built via explicit labeling or automatically via a set of heuristics (shared with users) for labeling email, based on email review activity. The probability distribution over urgency is used to compute a measure of the *expected cost of delayed review* [7]. The classifier considers multiple attributes of messages including organizational relationships between the recipient and senders, the proximity of the message composition time to key times and dates gleaned from messages, the length of messages, and multiple textual distinctions in the header and body of messages, including words, predefined phrases, the existence of questions, the use of past, present, and future tense, text patterns, such as capitalization of words, and Boolean combinations of features, such as the recognition of questions in messages from a manager or direct reports. A view of the desktop client for PRIORITIES is displayed in Figure 1. As displayed in the figure, the system provides a view of unread email, scoped by time and sorted by urgency. Beyond the desktop, PRIORITIES makes decisions about if and when to send messages to a user's mobile device, such as a pager, PDA, or cell phone.

PRIORITIES considers multiple features of a user's context to modulate desktop and mobile notifications about urgent messages. We integrated into PRIORITIES an application monitoring capability, harnessing the EVE event monitoring system [5], which enables the system to sense user activity.

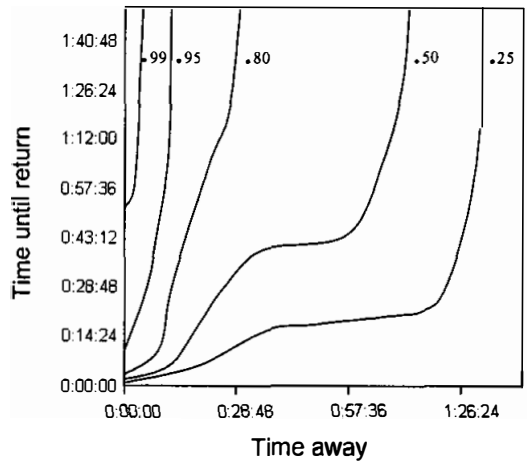

Figure 2. Forecasts for time until return at different probabilities (labeled), conditioned on time away.

PRIORITIES allows users to specify context-sensitive urgency thresholds for desktop and mobile alerting about incoming messages. In desktop settings, these thresholds are specified as functions of different kinds of desktop activity (*e.g.*, user has been typing within a predefined time horizon) and the status of an ambient acoustical analysis, configured to capture whether the user is having a conversation. The system also considers the user's meeting status as represented in their Outlook calendar. Users can set thresholds on the minimal urgency required for alerting for background *no-meeting* situations as well as for the cases where they are having a normal or a specially marked meeting. The system examines the current meeting status and may hold an alert about an urgent message until a meeting ends and the alerting threshold returns to the lower no-meeting urgency setting.

For mobile settings, PRIORITIES seeks to minimize the liklihood that a user, working at or near a desktop system, will receive messages on their mobile device. Incoming messages surpassing the context-sensitive threshold are relayed to a mobile device via an SMTP connection only if the system detects that the message has been waiting for at least some user-specified period of time, referred to as a *user-away* period. User-away periods are segments of time where no computer or conversational activity has been detected at the user's main desktop system.

Turning to the central focus of this paper, early versions of PRIORITIES considered preferences about prespecified, deterministic user-away periods required before messages would be sent to mobile devices. However, in later systems, we introduced a version containing a presence-forecasting component. This subsystem maintains a log of a user's presences, and can make mobile-messaging decisions based on identifying the probability distribution over the duration of the user-away period. Rather than waiting a fixed amount of time before sending messages, the presence-forecasting component forwards urgent messages to mobile devices when the user is likely to be away for some user specified period of time. Thus, urgent messages may be transmitted earlier than with use of the fixed policy.

PRIORITIES continues to collect data about a user's activity on a computer, as well as ambient acoustics near the computer. The system logs events and durations representing a user's time at and away from the computer. In addition to observing keyboard, mouse, and acoustic events, we also record the time of day and note the status of a user's calendar during the current period, using the Outlook Collaborative Data Objects (CDO) and Active Directory interfaces to access appointment information. PRIORITIES considers patterns of presence at different times of day and days of the week, by decomposing days into a set of prototypical time periods for weekdays and weekends, including durations of several hours representing *morning, lunchtime, afternoon, evening,* and *night*.

PRIORITIES issues queries to its growing database of presence and absence periods to generate cumulative probability distributions via direct conditioning on reference classes defined by automated queries. A reference class includes cases consistent with the current context, including a specification about the period of time a user has already been away so far, the time of day, day of week, and calendar status. The system outputs a cumulative distribution for the time a user will return, $p$(return by $t$ | time away, time of day, day of week, calendar status).

As part of PRIORITIES research on presence forecasting, we explored the influence of time away, time of day, and day of week on cumulative distributions for return to a desktop computer. Figure 2 shows a set of curves, built from the presence database of a PRIORITIES user, representing the times until the user will return to their main desktop system, given that they have been away for different periods of time. The figure shows distinct curves for different probabilities of arrival by the time indicated. Figure 3 displays smoothed curves, representing forecasts for the same user, of the likelihood of returning to a main desktop computing system within 15 minutes, given that he has been away for periods of time extending up to 40 minutes. The darker curve represents the forecast considering only the amount of time away. The dashed curves represent forecasts conditioned more tightly on data segmented into different prototypical regions of time, including *morning*, *lunchtime*, *afternoon*, *evening*, and *night*. As highlighted in the Figure 3, we discovered that considering the time of day is valuable in forecasting presence. Other studies showed the value of additionally conditioning on the day of week.

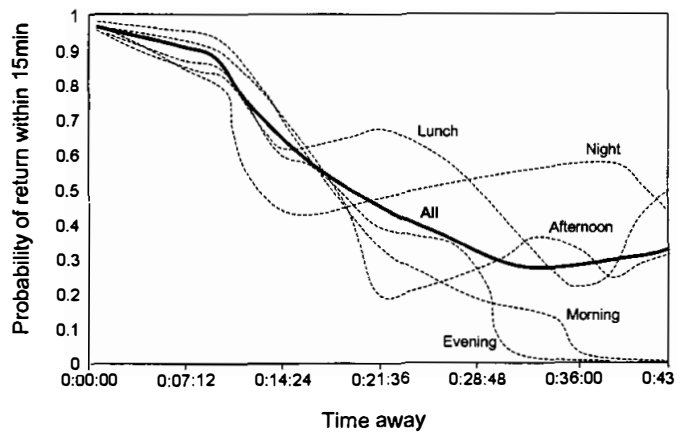

Figure 3. Probability of a user returning within 15 minutes given time away for all weekday data (bold), and for data segmented into prototypical segments of a weekday.

We found that there was enough data after several weeks of monitoring users to compose informative cumulative distributions about presence in an on-demand manner via direct conditioning over the PRIORITIES presence log. Nevertheless, we sought to develop methods for handling the potential problem of having inadequate data about particular queries. To address this challenge, we integrated a procedure that increases the number of cases considered as relevant by progressively broadening the reference class used to define cases as the quantity of data diminishes. With increasing amounts of data, the reference class is tightened into increasingly finer contexts, according to a heuristic sequencing of attributes that condition cases on increasingly precise details of the time of day, day of week, and meeting status.

In the deployed system, we worked to minimize the complexity of user controls and to make policies understandable to users. PRIORITIES allows users to specify a probability of being away for time $t$ as an assertion that the user is "away for at least $t$" for simplifying controls. For example, for the mobile alerting function, a user can specify that a probability of 0.85 or more of being gone for some prespecified time $t$ can be used to turn on the forwarding of urgent mobile messages immediately rather than waiting for an urgent message to age 30 minutes.

In summary, the presence-forecasting subsystem of PRIORITIES formulates predictions at run time about the

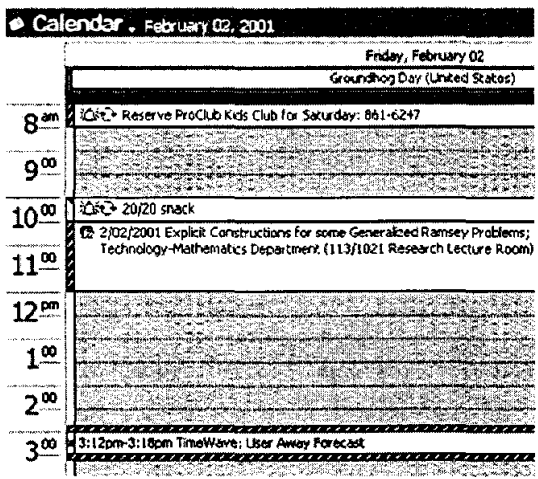

Figure 4. TIMEWAVE posting of a user-away forecast (shaded segment at bottom) on a shared calendar.

Probability distribution over periods of forthcoming absence for the appropriate reference class by accessing data for periods of absence conditioned on the amount of time away so far, the relevant time of day, and the day of the week.

### 2.2 TimeWave and SmartOOF Services

Beyond the use of presence forecasting for decisions about the timing of mobile alerts, we harnessed the user-away forecasts in two other services, named TIMEWAVE and SMARTOOF, which were integrated into PRIORITIES. Both services allow users to share with colleagues predictions about when they will return to their office if they are away. Users can select a probability of return that is taken as the time they are "likely to return" for the purposes of communication. As an example, a user can assert that the system should relay the time in minutes until they will return to the office with a 0.9 probability. Given such a preference, the presence forecasting system identifies the period of time until they will have returned with a 0.9 probability, given the time of day, and the time away so far.

The TIMEWAVE feature automatically populates a user's free-busy information with a specially marked *away forecast* when the user steps out of the office. Thus, TIMEWAVE allows users to share out this period of time as an away period that can be viewed by others who have access to their online calendar. Figure 4 displays a situation where a user has just stepped out of the office for several minutes in the afternoon. The system infers that the user will be back within six minutes with a 0.9 probability and posts this information on the shared Outlook calendar. TIMEWAVE owes its name to the notion that the user-away forecast is continually updated, leading to an ebb and flow of the predicted away period with changes in time away.

SMARTOOF allows users to encode preferences about the selective emission of *out-of-office* (OOF) messages based on the inferred urgency of incoming messages, and on the results of an availability forecast. A user can encode preferences to send the special out-of-office message if he or she is likely to be away for some specified time, given that an incoming message has at least some threshold value of urgency. The user can also tell the system to include an availability forecast within the OOF message.

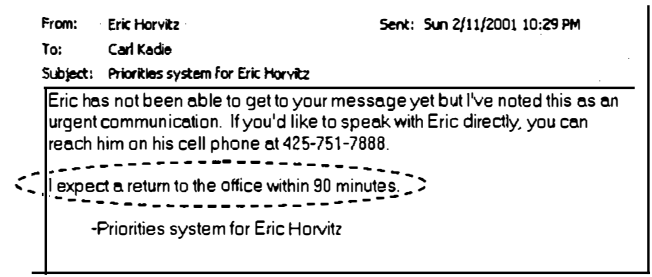

Figure 5. A message generated by SMARTOOF, relaying a presence forecast for the recipient to the sender of an urgent message when the recipient is likely to return to their office.

Figure 5 displays a typical message used by the SMARTOOF service. In this case, the availability forecast is used to communicate with a colleague that the recipient of an urgent message will likely return to their desktop system within 90 minutes.

## 3 Learning and Inference in Coordinate

The experience with presence forecasting in PRIORITIES stimulated us to pursue more general Bayesian presence forecasting machinery within the COORDINATE project. As the project matured, rather than rely on the progressive specialization of reference classes, we sought to employ Bayesian learning and inference to discover generalizations and to provide a means for fusing multiple distinctions about time, activity, and such rich contextual clues as details about numerous properties of appointments captured on a calendar. Bayesian learning and inference also provide a principled method for addressing the potential sparseness in data associated with early phases of data collection.

We also worked to extend the scope of the system's data collection and reasoning. One limitation of presence forecasting in PRIORITIES' is the reliance for modeling and prediction on the collection of data from a single machine. To build more general predictive models, we need to collect data about a user's activity and location from multiple sources, including data about a user's activities on multiple devices in addition to data from a calendar. We also sought to generalize forecasts about presence and absence to other events of interest to support collaboration and communication. For example, as we shall see in the discussion of applications in Section 6, we wish to understand if and when a user will access messages waiting in their inbox, or to identify a good future time to interrupt the user with a notification. We also would like to forecast when a user will have easy access to computing systems or devices with particular capabilities. For example, we might like to know when a user will likely have easy access to a computer with full videoconferencing abilities.

### 3.1 Coordinate Components

COORDINATE was built as a server-based service written in C# within the Microsoft .NET development environment. A schematized overview of the system is displayed in Figure 6. The prototype includes a central database, networking facilities, device provisioning interfaces, and Bayesian machine learning tools. The system was engineered primarily to serve as a facility for use by automated proxies that provide collaboration and communication services to users rather than to be queried directly by users. However, a query interface allows researchers to directly query the service.

COORDINATE is composed of four core components. The *data-acquisition* component executes on multiple computers that a user is likely to use. The component detects computer activity, calendar information, and video, acoustical, and position information from 802.11 wireless signal strength and GPS data when these channels are available. The data-acquisition component includes a signal-processing layer that allows users to configure parameters of audio and video sources used to define presence. The information is cached locally and sent to a COORDINATE *data-coalescence* component running on a central COORDINATE server. This component is responsible for combining the data from the user's multiple machines and storing it in an XML-encoded event database.

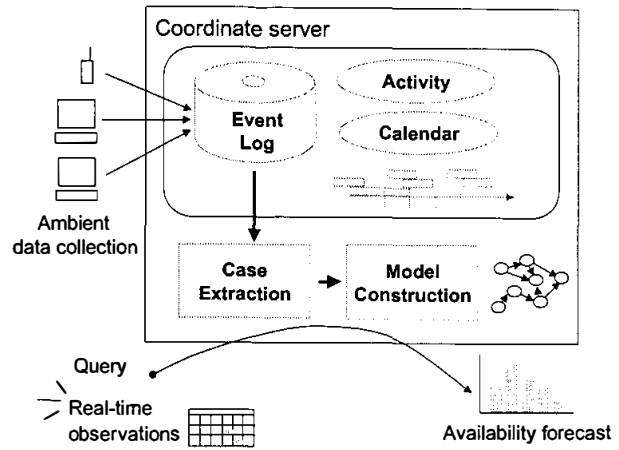

Figure 6. Schematized overview of COORDINATE.

Rather than attempting to build a massive static predictive model for all possible queries, we instead focus the analysis by constructing a set of cases from the event database that is consistent with the query at hand. This approach allows us to custom-tailor the formulation and discretization of variables representing specific temporal relationships among such landmarks as transitions between periods of absence and presence and appointment start and end times, as defined by the query. A set of cases is fed to the fourth component, a *learning and inference* subsystem, which constructs one or more Bayesian networks that are custom-tailored for the target prediction. The Bayesian network models are used to compute cumulative distributions over events of interest. We employ the WinMine learning tool, developed by Chickering, Heckerman, and Meek [3], to perform structure search over a space of dependency models, guided by a Bayesian model score to identify graphical models with the best ability to predict the data.

### 3.2 Coordinate Analyses

COORDINATE logs periods of presence and absence in a manner similar to logging in the earlier PRIORITIES effort. However, in COORDINATE, events are annotated by the source device. Descriptions of devices, including capabilities and location are maintained in a devices profile. For example, a user can specify that a device is based in a user's office and has full videoconferencing abilities. The tagging of events by specific devices, indexed into locations and capabilities allows the system to forecast a probability distribution over the time until the user will have access to different kinds of devices. When these devices are assigned to fixed positions, forecasts can be made about location.

COORDINATE's event system can monitor the history of a user's interaction with computers, including applications that are running on a system, applications that are now in focus, and those that have just gone out of focus. As an example, the system can identify when a user is checking email or reviewing a notification. Thus, moving beyond presence and absence, COORDINATE supports such forecasts as the time until a user will engage an application or cease using an application. Thus, the system can be queried about when a user will likely access his or her email inbox to review new messages. As the system also detects conversations, we have been experimenting with its ability to predict when a proximal conversation is likely to end.

COORDINATE provides forecasts $p(t_e \mid E, \xi)$, where $t_e$ is the time until an event of interest, $\xi$ is the background state of information, and evidence $E$ includes *proximal activity context*, the time of day, day of week, multiple attributes representing the nature of active calendar items under consideration, and other observations that can be added to consideration (*e.g.*, desktop activity). The proximal activity context represents one or more salient recent transitions among landmark states, based on the query. Such conditioning captures a modeling assumption that times until future states are strongly dependent on the timing of the most recent key landmarks. For predictions about the time until a user who has been absent will return to their office, or return to their office and remain for at least some time $t$, the proximal activity context is the period of time since the user transitioned from present to absent. For forecasts about how long it will be before a user who is present will leave their office, or, more specifically, will be away for at least some time $t$, the proximal activity context is taken as the time since the user transitioned from absent to present. In response to a query, COORDINATE's case-acquisition component identifies a set of cases that fit the proximal activity context defined by the query, and associates the cases with other concurrent observations (*e.g.*, day of week, time of day, appointment properties), and constructs Bayesian networks used to make predictions.

Figures 7 displays COORDINATE's research interface that provides a means for selecting classes of queries and formulating queries for real-time or offline analyses. In the case pictured, a query has been entered about the likelihood that a user will return to the office for a period of at least 15 minutes, given that he has been absent for 25 minutes at 10:15 AM on a weekday. Relevant cases are gleaned from the event database and a Bayesian network is learned. The network is used to generate the displayed cumulative probability distribution about when the user will return. The system also shares a text summary forecast based on a predefined confidence threshold (in this case 0.8). Similar analyses are performed for other events of interest. In Section 5, we describe methods for folding in consideration of meetings and the construction of models that provide forecasts of a user's availability. First, we shall review the construction of models of meeting attendance, interruptability, and location.

## 4 Models of Attendance and Interruptability

A subproject of the COORDINATE effort focuses on methods for analyzing properties of meetings and integrating these observations into the overall analysis of presence and availability. Beyond enhancing predictive models about presence, we can learn models that relate attributes of an appointment to the likelihood that a meeting will be attended, and to the interruptability of meetings. We can also learn models for inferring the location of meetings when locations are not clear from location fields. Such inferences can provide useful inputs for shared calendars, as well as a set of valuable parameters for automated notification and collaboration systems.

### 4.1 Learning about Attendance and Interruptability

COORDINATE logs all meetings stored in a user's calendar, noting the status of properties of appointments noting the status of properties of appointments made available in the online Outlook calendar in addition to several additional computed properties. The data is used to learn models that can predict attendance, interruptability, and location.

Prior work on probabilistic models for predicting a user's attendance at meetings has relied on handcrafted models. In research on the Attentional User Interface system [7], a Bayesian network was constructed to consider several properties of appointments to predict a user's location, availability, and attention, based on appointment properties. In other work, Mynatt and Tullio [9] describe construction of a Bayesian network by hand that can be used to estimate the likelihood that a user will attend a meeting from multiple properties of meetings and setting.

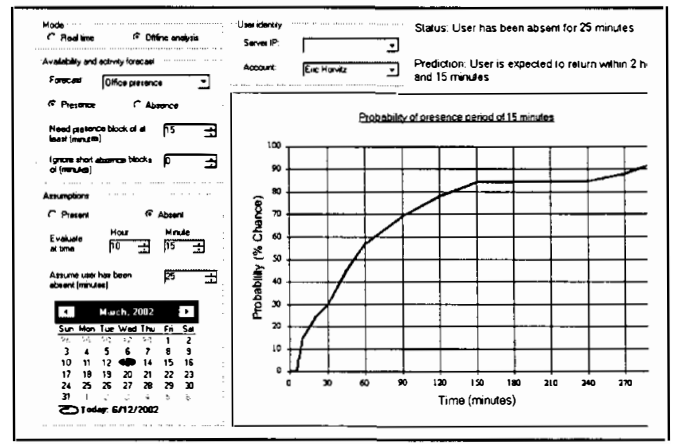

Figure 7. Query interface for COORDINATE showing a cumulative probability distribution answering a query about when a user will return to his primary office for at least a 15 minute block of time, after an absence of 25 minutes in the morning on a weekday, when no meetings are scheduled.

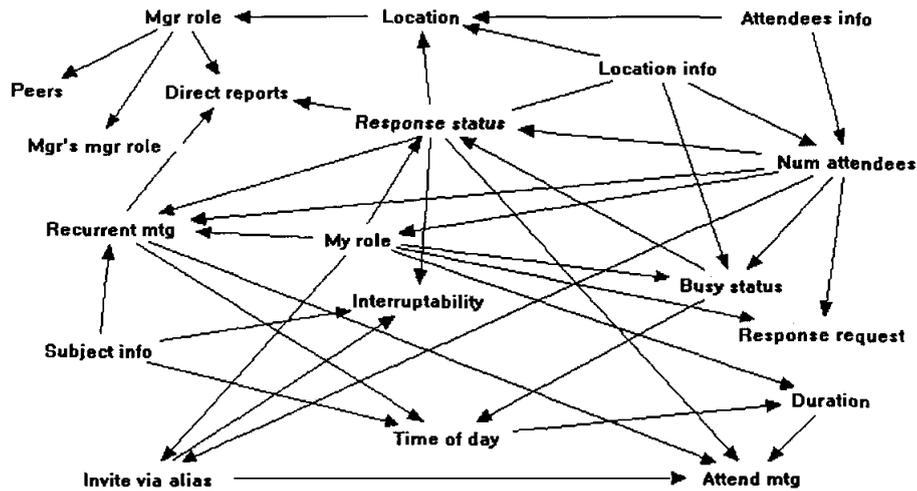

Figure 8. Bayesian network learned from online data that predicts the likelihood of attending meetings, and probability distributions over the interruptability and location of the meetings.

For building models of attendance, COORDINATE creates a draft training set of appointments and their properties, and marks an attendance field for each appointment with guesses about attendance made via the use of a set of heuristics about attendance based on monitored activity. The attendance heuristics consider the sensing of desktop activity through a significant portion of a scheduled meeting as evidence that a meeting was not attended and, likewise, the lack of activity during large portions of a meeting as evidence that a meeting was attended. As the activity-based heuristics for annotating meeting attendance provide noisy labeling, we take COORDINATE's guesses about attendance as rough guesses about attendance. The system provides a form that allows users to refine the draft with manual labeling of attendance. Beyond editing the attendance field of appointments, users also can add assessments of the physical location of meetings, and how interruptible they are during the different meetings.

COORDINATE can generate a form that lists meetings in order of their occurrence and displays an attendance field containing the different meetings, specifying whether meetings are *low*, *medium*, or *high* interruptability. The annotated meeting log is used to construct a model that can predict the likelihood that a user will attend a meeting, given appointment properties. Beyond time of day and day of week, meeting properties considered at training and prediction time include the meeting date and time, meeting duration, subject, location, organizer, number and nature of the invitees, *role* of the user (user was the organizer versus a required or optional invitee), response status of the user (responded yes, responded as tentative, did not respond, or no response request was made), whether the meeting is recurrent or not, and whether the time is marked as busy or free on the user's calendar. The system accesses the Microsoft Active Directory service to recognize and annotate organizational relationships among the user, the organizer, and the invitees, noting for example, whether the organizer and attendees are organizational peers, direct reports, managers, or managers of the user's manager.

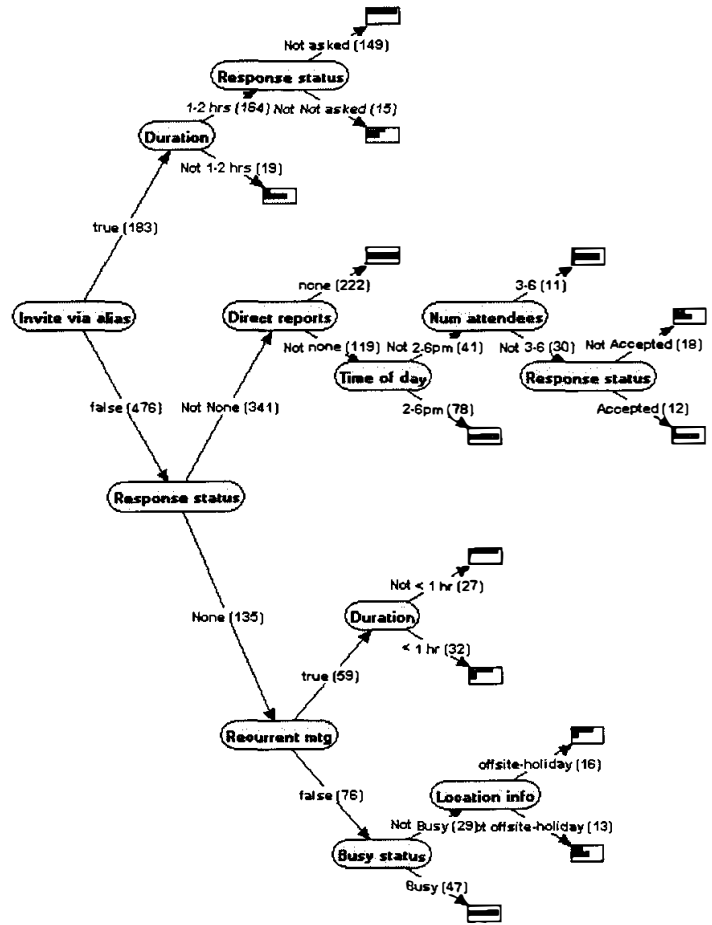

Figure 9. Decision tree for predicting the probability that a user will attend a meeting, based on properties of the meeting encoded in an appointment item.

We performed several experiments as part of an evaluation of the accuracy of the predictive model for calendar attendance, interruptability, and location. Figure 8 displays a Bayesian network learned from the data of a single user that shows the probabilistic dependencies among variables extracted from an online calendar and variables of interest represent. The model was constructed by collecting data from a six-month period of meetings stored in a user's online calendar between October 2001 and March 2002. The data set includes 659 appointments. The initial 559 appointments were used to train the model, leaving the last 100 cases as a holdout set for testing. The owner of the calendar was asked to annotate cases with information on whether meetings were attended, to note the location of the meetings, and also to indicate the interruptability of the meetings, discretized into low, medium, and high interruptability. For the data set, 0.64 of the appointments were attended. The user assigned 0.5 of the cases an interruptability property of low, 0.4 of the cases medium, and 0.1 of the cases high. The model performs well; the classification accuracy on holdout data was 0.92 for predicting attendance and 0.81 for assigning interruptability.

Decision trees for predicting meeting attendance and interruptibility are displayed in Figures 9 and 10 respectively. As displayed in Figure 9, key influencing variables for predicting meeting attendance include whether the meeting is organized via a mailing list (referred to as a group *alias*) or by an individual, the duration of the meeting, the response status, whether the meeting is recurrent or not, the number of attendees, whether the user's direct reports have been invited, the information included in the location field, and whether the meeting is marked as busy time or not. The bar graphs at the leaves of the decision tree display the probability of attendance versus non-attendance, with the event $p$(not attend $| E,\xi$) at the top position, followed by $p$(attend $| E,\xi$). As indicated in Figure 10, the main influencing variables for predicting the interruptability of meetings is whether a user organized by a group *alias* versus an individual, whether the user responded to the appointment, the number of attendees, whether the user's direct reports are invited, and the subject of the meeting. The probability distributions over interruptability are displayed as bar graphs at the leaves of the decision tree where the states from top to bottom are low, medium, and high interruptability.

## 5 Integrating Attendance and Interruptability

COORDINATE employs models of attendance and interruption in several ways. The system allows direct queries about the probability that a user is attending or will attend a meeting.

### 5.1 Expected Cost of Interruption

COORDINATE also can share information about the *expected cost of interruption* (ECI) for a user at the present moment

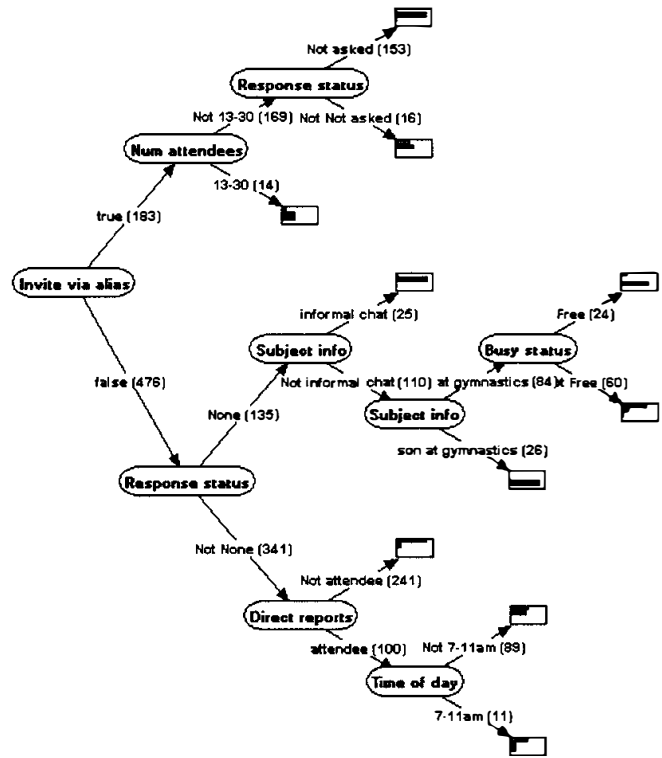

Figure 10. Decision tree for predicting the interruptability of meetings constructed from training data.

or at future times. Users are provided with a facility to associate a dollar-value cost of interruption for each interruptability level. They also can assess default costs of interruption for free periods for prototypical times of day and days of week. The system computes an expectation,

ECI=

$$p(A^m \mid E,\xi)\sum_i p(c_i^m \mid E,\xi)c_i^m + (1 - p(A^m \mid E,\xi))c^d$$

where $A^m$ is the event of attending a meeting, $c_i^m$ is the cost of interruption associated with interruptability value $i$, $c^d$ is the default cost for the time period under consideration, and $E$ represents appointment attributes, the proximal context, time of day, and day of week. Such an expectation can be provided to authorized colleagues who wish to identify a good time to initiate a communication. The cost of disruption can also be used to inform cost–benefit analyses in automated notification and communication systems, such as the BESTCOM service described in Section 6.

### 5.2 Considering Meetings in Presence Forecasting

COORDINATE integrates inference about the nature and timing of meetings into its predictions about absence and presence. The system performs an approximate meeting analysis to forego the complexity of considering multiple patterns of meetings. In the approximation, we make an assumption of meeting independence, and consider each meeting separately. A subset of meetings on the user's calendar are considered to be *active* for the query based on their proximity to the times and transitions dictated in the query. For each active meeting, a distinct Bayesian network model and associated cumulative distribution is computed for the time until presence or absence over the course of the meeting's *scope*—a period of time created by adding predefined lengths of time before and after the explicit meeting boundaries. In constructing the model for each meeting, the case-acquisition component of COORDINATE identifies cases that are consistent with the proximal context defined by the query. Only meetings that had been marked as attended are considered in this phase.

In the last step, the cumulative distribution for the time until return over the scope of each meeting is combined with the cumulative distribution for the no-meeting situation for that period of time. The system performs this combination by first constructing a cumulative distribution for a presence transition for the *no-meeting* situation. This cumulative distribution is computed in the manner described in Section 3, employing cases consistent with the query where no meetings were scheduled or where the user indicated that a meeting was not attended. Then, for the span of time represented by each meeting's scope, the cumulative distributions for the *attend* and *no-attend* situations are summed together, weighted by the inferred likelihoods that the user will and will not attend the meeting, respectively.

Figure 11 shows the influence of the integration of inferences about the likelihood of attending meetings on the forecast of a user's availability. A query has been submitted at 1:20 PM on a weekday seeking a forecast about the period of time until a user is expected to return to their desktop machine when they have already been absent for 15 minutes. The gray curve shows the cumulative distribution for the time until the user will return for the no-meeting situation. The black curve shows the result of folding in a consideration of active meetings, taking into account the likelihood that the user will attend each meeting. In this case, three meetings have been considered, including appointments at 1:00-2:00 PM, 2:30-3:30 PM, and 4:00-5:00 PM. Bayesian network models are constructed automatically for each of these meetings.

## 6 Notification Platform and Bestcom

COORDINATE was designed primarily to support NOTIFICATION PLATFORM and BESTCOM, two communication services developed at Microsoft Research. NOTIFICATION PLATFORM is a general notification architecture that provides interfaces and subscription mechanisms for integrating multiple sources of notifications. Interfaces are also provided for provisioning multiple devices. A *notification manager* considers the urgency of messages and the cost of disrupting users based on contextual information that is provided by a *context server* component. The context server has access to presence and availability information, and to encodings of a user's preferences about the value of message information and the cost of interruption in different settings. In decision-theoretic variants of the notification manager, a probability distribution over the time until a user will review a message, in the absence of an explicit notification, is used in computing the expected value of relaying an alert. COORDINATE provides probability distributions over time until the review of information as well as contextual information about the expected cost of interruption that is consumed by the notification manager.

COORDINATE's inferences and query classes have been shaped by the needs of BESTCOM, a descendant of NOTIFICATION PLATFORM, that centers on automated and semi-automated mediation of interpersonal communications. BESTCOM provides people with best-effort communications based on a consideration of context, available channels and preferences about communications. In BESTCOM, a communications decision-making agent acts as a proxy, and considers the goals and context of a contactor and contactee. Although preferences of both participants of a communication may be considered, incarnations of BESTCOM typically put strong weight on the preferences of the contactee, as it is the contactor who typically seeks attentional resources from the contactee.

BESTCOM considers preferences of contactees about how to handle incoming communications, based on the identity of the contactor, the initial channel selected (*e.g.*, telephone, instant messaging, email), and the inferred or annotated goal of the contactee. In situations where the BESTCOM service is explicitly invoked by the contactor, annotations about the nature of the communication may be shared as part of a communications metadata schema. For example, a contactee may wish to invoke BESTCOM to speak with a coauthor about a specific change she wishes to make to a shared document by invoking BESTCOM within a particular location in a word-processing application. A BESTCOM

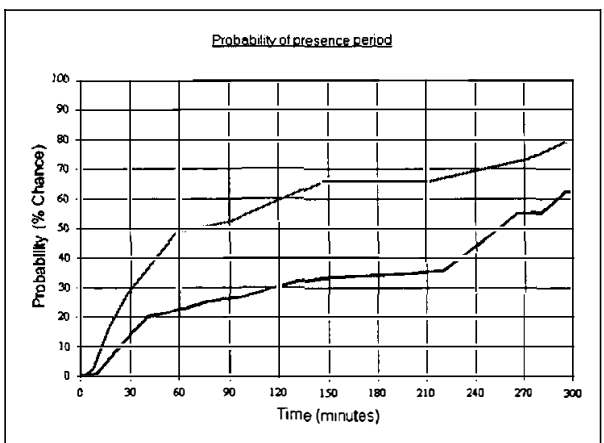

Figure 11. Influence of meeting analysis on a presence forecast. Upper curve shows the cumulative distribution for the no-meeting situation. Lower curve integrates likelihoods of attending meetings found on the user's schedule.

service can share the contactor's goals and available channels with the contactee's proxy.

The spirit of BESTCOM is to maintain privacy about a contactee's state. Although contextual information is used in deliberating about the situation, the contactor is typically aware only of summary decisions about how to handle the communication. Actions include establishing a real-time connection on the same channel, shifting to another channel, taking a message, and providing the contactor with better times for communicating with the contactee, coupled with services to schedule and manage the future communication.

BESTCOM efforts have included the development of decision models and preference-assessment tools that provide such facilities as efficient means for creating and editing groups of people that users wish to assign different communication priorities, and assessing the cost of interruption in different contexts. Some versions of BESTCOM leverage direct assessments of preferences about people and context, and rely on a direct sensing of contactee's state. More sophisticated versions rely on richer contextual inference, such as forecasts about presence and availability provided by COORDINATE. In decision-theoretic approaches to BESTCOM, a contactee's *communications manager* considers the expected cost of interruption associated with accepting incoming communications and reasons about the change in the expected value of the interaction for such potential actions as deferring the communication or shifting communication channels. Learning and inference for forecasting a user's presence, availability, and such states as location and device access under uncertainty, provide BESTCOM services with valuable information for decision making about incoming communications.

## 7 Summary

We reviewed research on the challenge of forecasting computer users' presence and availability. We first reviewed an earlier forecasting subsystem that was built as an embedded component of the PRIORITIES prototype. The subsystem was used to extend the system's mobile messaging abilities, as well as to provide new services, including SMARTOOF, which provides selective out-of-office messaging, and TIMEWAVE which shares a user's future presence via automated updating of a shared calendar. We then described the creation of a more general presence and availability forecasting service, named COORDINATE, with the ability to log events from multiple devices. We reviewed the learning of Bayesian models from log data for forecasting states of interest. After reviewing basic functionalities with regards to forecasting presence, we described the learning of Bayesian networks to predict meeting attendance and interruptability, and the integration of a consideration of meetings into presence forecasts. Finally, we reviewed the value of COORDINATE inferences in support of decision making in the NOTIFICATION PLATFORM and BESTCOM messaging and communication services.

Research on the COORDINATE project is ongoing. Our experiences with the prototype COORDINATE service have excited us about the promise of building richer situation forecasting tools and services. They have also heightened our awareness of the challenges associated with the effort required for data collection and labeling that we need to address on the path to general deployment of the system. To date, the system has been fielded only within our research team. We are working towards refining the system and fielding and testing future versions of COORDINATE in a larger group setting.

### Acknowledgments

We are grateful for the assistance provided by David Hovel with the EVE eventing system, PRIORITIES, and with predecessors of components of COORDINATE.